\documentclass[onecolumn]{emulateapj}
\usepackage{amssymb,bm,graphicx,natbib}

\renewcommand{\k}{\ensuremath{\bm{k}}}
\renewcommand{\v}{\ensuremath{\bm{v}}}

\begin{document}
\title{Large-scale BAO signatures of the smallest galaxies}
\author{Neal Dalal}
\author{Ue-Li Pen}
\affiliation{Canadian Institute for Theoretical Astrophyics,
  University of Toronto, 60 St.\ George St., Toronto, Ontario M5S 3H8,
  Canada} 
\author{Uros Seljak}
\affiliation{Physics Department and Lawrence Berkeley National
  Laboratory, University of California, Berkeley, California 94720, USA}
\affiliation{Institute for Theoretical Physics, University of Zurich,
  Zurich, Switzerland} 
\affiliation{Institute for Early Universe, Ewha University, Seoul
  120-750, S. Korea} 

%\maketitle

\begin{abstract}
Recent work has shown that at high redshift, the relative velocity
between dark matter and baryonic gas is typically supersonic. This
relative velocity suppresses the formation of the earliest baryonic
structures like minihalos, and the suppression is modulated on large
scales. This effect imprints a characteristic shape in the clustering
power spectrum of the earliest structures, with significant power on
$\sim$100 Mpc scales featuring highly pronounced baryon acoustic
oscillations. The amplitude of these oscillations is 
orders of magnitude larger at $z\sim 20$ than previously expected.
This characteristic signature can allow us to
distinguish the effects of minihalos on intergalactic gas at times
preceding and during reionization. We illustrate this effect with the
example of 21 cm emission and absorption from redshifts during and
before reionization. This effect can potentially allow us to probe
physics on kpc scales using observations on 100 Mpc scales. 

We present sensitivity forecasts for FAST and Arecibo.  Depending
on parameters, this enhanced structure may be detectable by Arecibo at
$z\sim 15-20$, and with appropriate instrumentation FAST could
measure the BAO power spectrum with high precision. In
principle, this effect could also pose a serious challenge for efforts
to constrain dark energy using observations of the BAO feature at low
redshift.  
\end{abstract}

\section{Introduction}

Structure formation in the standard inflationary $\Lambda$-Cold Dark
Matter ($\Lambda$CDM) cosmological model is expected to proceed
hierarchically. The earliest bound, virialized structures arise on
small scales, and as time progresses these objects merge and accrete
mass, growing ever larger until cosmic acceleration at low
redshift freezes out the growth of large-scale structure.

The role played by the earliest generations of collapsed structures in
the thermal history of the universe is, at present, unclear.
Observations of the spectra of high-redshift quasars indicate that the
reionization of the intergalactic medium was largely complete by
redshift $z\approx 7$ \citep[e.g.][]{Fan06}, while measurements of the
Thomson scattering optical depth of the cosmic microwave background
suggest that reionization occurred at $z\sim 10$ \citep{WMAP7}. At
these epochs, the typical masses of collapsed dark matter halos range
from rare $10^9 M_\odot$ objects, down to (plausibly) Earth-mass halos
\citep{Green05}.  As we discuss below, the smallest dark matter halos
are unable to attract baryons, resulting in an effective lower mass
limit near $10^5 M_\odot$.

Even if low-mass halos are able to acquire baryons, they may be unable
to convert those baryons into stars \citep{Teg97,Bromm09,Loeb10},
since star formation requires the presence of cold, dense gas.
Objects massive enough to attract gas but whose virial temperatures
are below $\sim 10^4$ K, termed minihalos, cannot cool their gas
through atomic lines and must therefore rely upon molecular cooling
processes.  It is unclear whether molecular processes can cool
minihalo gas sufficiently to allow star formation.  Prior to the
formation of the first stars in the Universe, the formation of
molecular H$_2$ catalyzed by residual free electrons left over after
recombination appears insufficient to allow efficient cooling at
redshifts $z\lesssim20$ \citep{Teg97}.  However, feedback from the
first luminous objects can change this result.  Positive feedback, for
example from ionizing X-rays that strip electrons from atoms and
thereby spur molecule creation, could lead to efficient cooling.
Conversely, negative feedback in the form of ultraviolet
radiation in the Lyman and Werner bands could destroy molecules and 
suppress H$_2$ cooling over large volumes \citep{Yoshida07}.  
Given this wide range of
possible scenarios, it is unclear whether minihalos can form stars and
whether they might be important during the reionization of the
intergalactic medium.  This uncertainty, however, may be viewed as an
opportunity: any probe that can quantify the importance
of minihalos during redshifts preceding and during reionization would
dramatically help to elucidate the physics of star formation in the
first structures that arise in the Universe.

Recently, \citet[hereafter TH]{Tsel10} pointed out an important effect
governing the formation of $\sim 10^5 M_\odot$ minihalos, that had
previously been overlooked.  As we discuss below, this effect can
provide a minihalo signature in many potential observables.
TH noticed that the relative velocity
between dark matter and baryons following recombination is typically
supersonic. This relative velocity arises because dark matter is
accelerated by gravitational potential gradients, while baryons are
Jeans stabilized against gravitational collapse due to their tight
coupling with the photon radiation field, until recombination.
Because DM and baryons suffer different accelerations, they acquire
significant relative velocities that are predominantly sourced by
potential fluctuations on scales of order the sound horizon at
recombination, $\sim 150$ Mpc. At recombination, the baryon sound
speed and Jeans length fall precipitously, allowing baryons to respond
to the same gravitational potential wells that accelerate dark
matter. Subsequently, the large-scale relative velocity is unsourced
and decays as $a^{-1}$ following recombination. At redshifts $z\approx
1000$, the relative DM-baryon motion is highly supersonic, with Mach
numbers ${\cal M} \approx 5$. The gas sound speed does not initially
decay as quickly as $a^{-1}$ because of residual thermal coupling to
the CMB, so the Mach number diminishes over time to ${\cal M}\sim 2$
at $z\sim 100$, and remains nearly constant thereafter.

Because the relative motions between baryons and DM are supersonic,
they have significant effects on the growth of structure. TH computed
the effects of relative DM-baryon velocities at high redshift, in the
perturbative regime of structure formation. They showed that these
motions cause a $\sim 10\%$ suppression of the matter power spectrum
at $k\approx 200 {\rm Mpc}^{-1}$ compared to standard linear
perturbation theory calculations. Using a Press-Schechter approach,
TH also suggested that the abundance of dark matter halos of mass
$M\sim 10^6 M_\odot$ could be suppressed by a factor of $\sim 2$ at
redshift $z=40$. At later
redshifts closer to reionization, $z\approx 10-20$, the halo abundance
would be much closer to standard predictions.

In this paper, we consider a similar effect of the supersonic
relative velocities that TH discussed. Instead of considering the
effect on dark matter halos, we examine the impact on baryonic
objects, which likely determine the properties of observable
quantities like 21 cm absorption, emission, etc. The supersonic
flow changes the mass threshold of baryonically populated
halos, and this effect can be exponentially large for rare
objects.  The relative bulk flows between DM and baryons
are modulated on large scales, of order $\sim 100$ Mpc, meaning that
the minihalo abundance is similarly modulated.  This provides a
large-scale signature of the effect of minihalos.

All calculations presented here assume WMAP \citep{WMAP7} cosmological
parameters: $\Omega_m=0.27$, $\Omega_\Lambda=0.73$, $\Omega_b=0.045$,
$h=0.7$, $n_s=0.96$, $\sigma_8=0.8$.

\section{Baryonic collapse fraction}
\label{sec:fc}

In this section, we compute the statistics of the collapsed baryonic
density, which we parametrize using $f_c$, the collapsed fraction of
baryons. The baryonic collapsed fraction is different than the CDM
collapsed fraction because baryons, unlike CDM, have a nonzero
temperature, and therefore cannot fall into shallow potential wells.
Naively, we might expect that the characteristic mass $M_c$ unable
to attract baryons would scale like the Jeans mass 
$M_J \propto c_s^3/GH$, however
cosmological hydrodynamic simulations \citep{Gnedin00,Naoz10} have
found instead that $M_c$ scales like the filter scale
\citep{Gnedin98}, 
\begin{equation}
M_F^{2/3} = \frac{3}{a} \int_0^a da^\prime M_J^{2/3}(a^\prime)
\left[1-\left(\frac{a^\prime}{a}\right)^{1/2}\right].
\label{Mc}
\end{equation}
For a standard WMAP cosmology, 
$M_c\sim 2-3\times 10^4 M_\odot$ at $z\sim 20$ \citep{Naoz10}. 

Just as gas pressure impedes the collapse of baryons, we expect that
bulk relative velocity between gas and dark matter halos will also
suppress the accretion of baryons.  We can make a simple estimate of
the minimum halo mass that can accrete gas moving at some bulk
velocity $v$ relative to the halo through analogy to the above
argument.  Assuming that bulk kinetic energy is
converted into thermal energy when gas falls into the halo, we expect
that Eqn.\ (\ref{Mc}) will still hold if we replace the sound speed
$c_s$ with $c_{s,{\rm eff}}=(c_s^2+v^2)^{1/2}$.  Since $c_s$ and $v$
both scale as $a^{-1}$ (see below), this effectively multiplies the
critical mass scale by the factor $(1+v^2/c_s^2)^{3/2}$.

Given $M_c$, we can compute the baryon collapsed fraction $f_c$ by
integrating the halo mass function,
\begin{equation}
f_c={\bar\rho}^{-1}\int_{M_c}^\infty M\frac{dn}{dM} dM,
\label{eqn:fc}
\end{equation}
where we use the fitting function of \citet{ShethTormen} to describe
the halo mass function $dn/dM$.
Since $M_c$ is a function of the local relative velocity between CDM
and baryons, $\v_{cb}\equiv\v_c-\v_b$, we see that
the collapse fraction $f_c$ is also modulated by this velocity,
$f_c(\bm{x})=f_c(v_{cb}(\bm{x}))$.  We have assumed that the velocity
dependence of the collapse fraction is a sharp cut-off in the mass
function, whereas hydrodynamic simulations
\citep[e.g.][]{Gnedin00,Naoz10} find a smooth transition in baryon
content of halos below and above $M_c$.  This distinction will not be
significant for our results: the important point is that the collapse
fraction is now a function of the local relative bulk velocity between
dark matter and baryons.  

\begin{figure}
\centerline{
\includegraphics[width=0.45\textwidth]{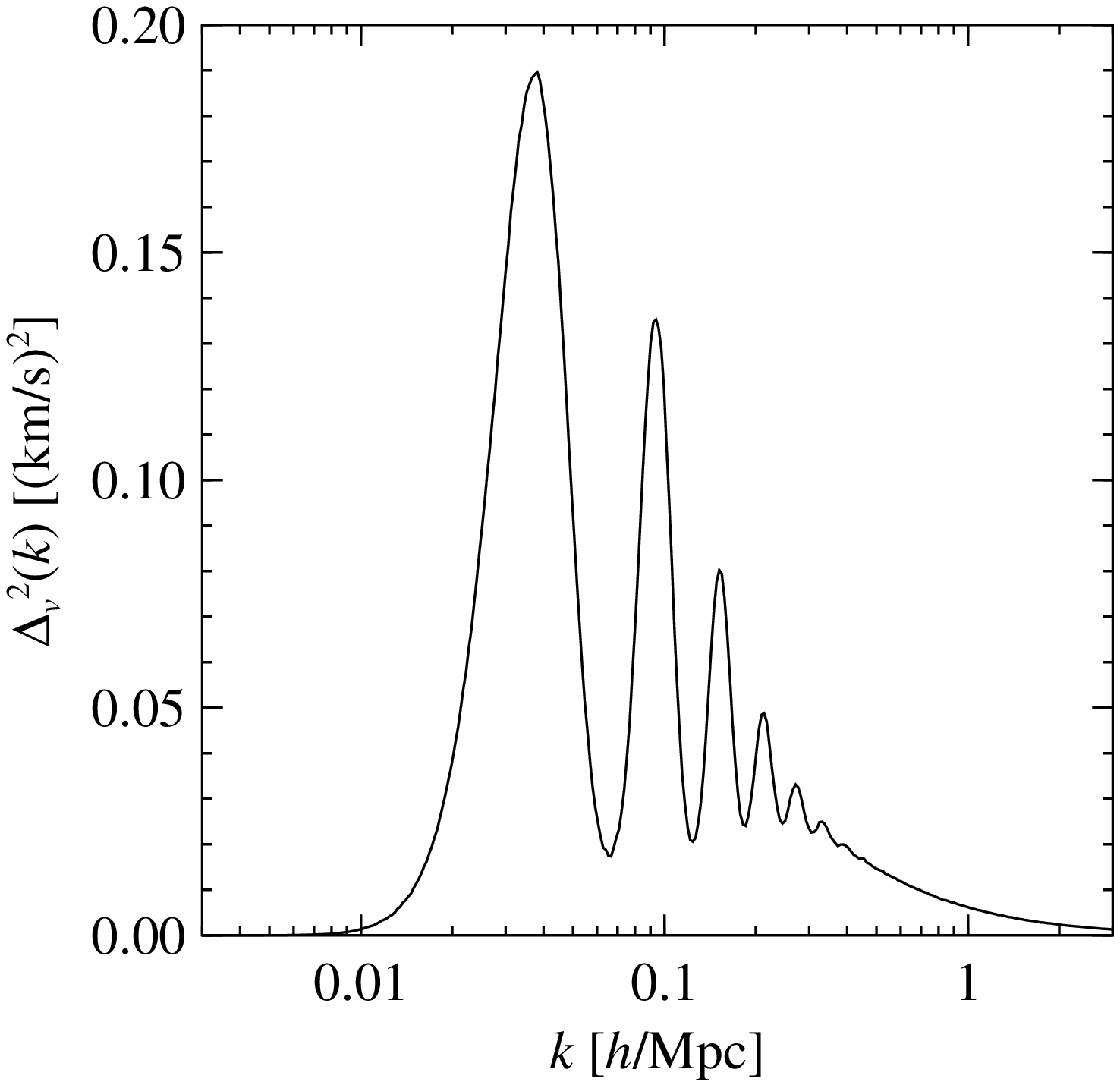} \qquad
\includegraphics[width=0.45\textwidth]{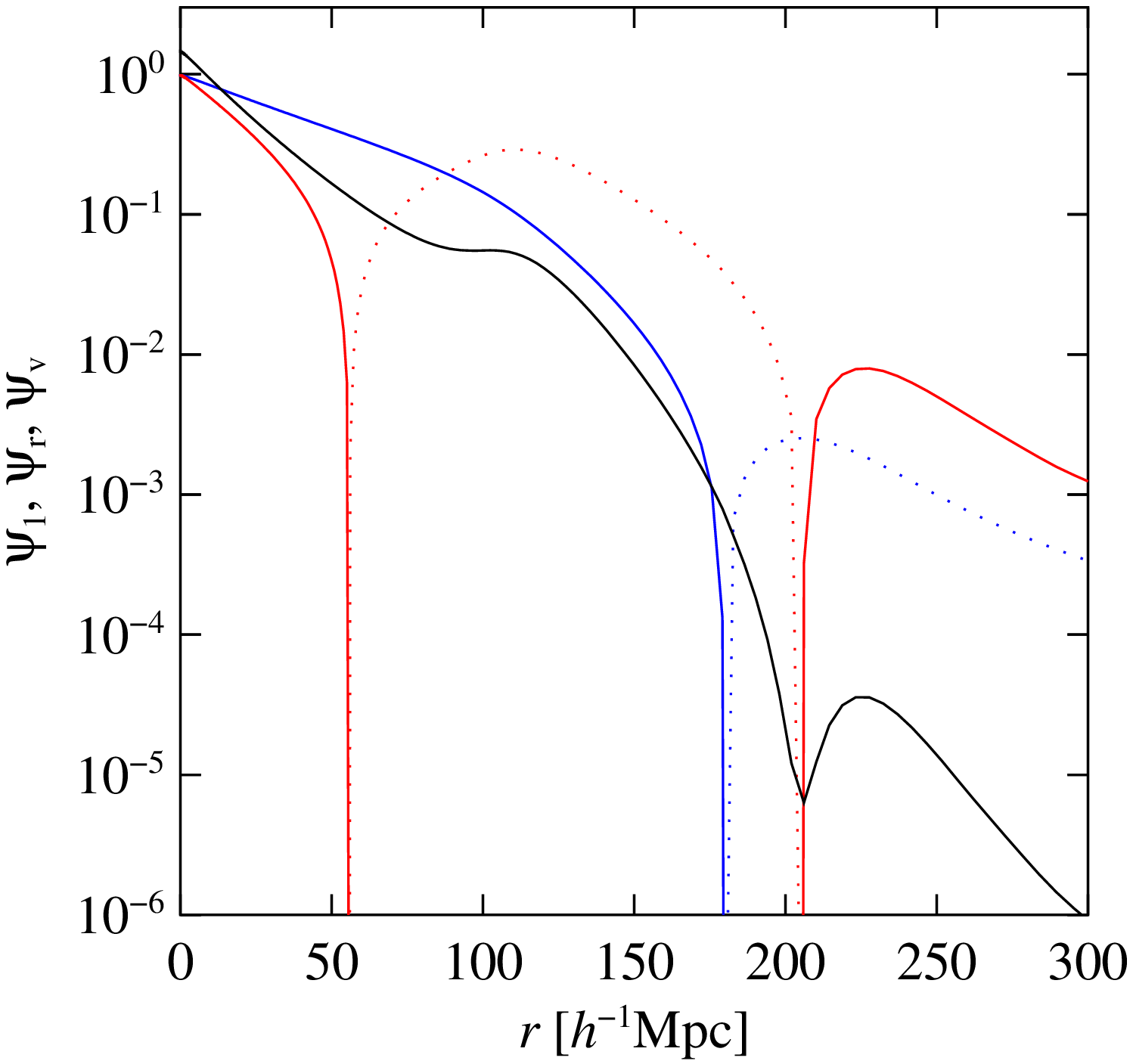}
}
\caption{(Left) The power spectrum of $v_{cb}$ fluctuations, 
$\Delta_v^2(k)=P_{\rm pri}(k) [a\,{\dot T}(k)]^2/2\pi^2k$,
at $z=15$. (Right) Velocity correlation functions
$\psi_1$ (blue), $\psi_r=\psi_1+\psi_2$ (red), and $\psi_{v}=\psi_1^2 +
\psi_r^2/2$ (black). Dotted curves depict regions where the
correlation function is negative.
\label{psi}}
\end{figure}

The relative velocity between CDM and baryons can be computed using
the linearized continuity equation
\begin{equation}
{\dot\delta} + a^{-1}\nabla\cdot\v=0,
\end{equation}
using comoving coordinates rather than proper coordinates. In the
linear regime, where we have potential flow, the velocity is
\begin{eqnarray}
\v(\k,a) &=& -i\frac{a\k}{k^2}{\dot\delta}(\k,a) \nonumber \\
&=&-i\frac{a\k}{k^2}{\dot T}(\k,a) \delta_{\rm pri}(\k).
\end{eqnarray}
For adiabatic perturbations, CDM and baryons have the same primordial
overdensity perturbations $\delta_{\rm pri}$, so the relative velocity
is
\begin{equation}
\v_{cb}(\k,a) = -i\frac{a\k}{k^2}{\dot T_{cb}}(\k,a) \delta_{\rm pri}(\k),
\end{equation}
where $T_{cb}=T_c - T_b$ is the difference between the CDM and baryon
transfer functions. The two-point correlation function for velocities
is then 
\begin{equation}
\langle v_i(\bm{x}) v_j(\bm{x}+\bm{r})\rangle = \sigma_1^2
\left(\psi_1(r) \delta_{ij} + \psi_2(r) \frac{r_i r_j}{r^2}\right)
\label{twopt}
\end{equation}
where
\begin{eqnarray}
\sigma_1^2 &=& \frac{1}{3}\int\frac{P_{\rm pri}(k)[a{\dot T}]^2}{2\pi^2}dk \\ 
\psi_1(r) &=& \frac{1}{\sigma_1^2}
\int\frac{P_{\rm pri}(k)[a{\dot T}]^2}{2\pi^2}\frac{j_1(kr)}{kr}dk \\
\psi_2(r) &=& -\frac{1}{\sigma_1^2}
\int\frac{P_{\rm pri}(k)[a{\dot T}]^2}{2\pi^2}j_2(kr).
\end{eqnarray}
Here, $P_{\rm pri}(k)$ is the primordial density power spectrum,
$\langle\delta_{\rm pri}(\k_1)\delta_{\rm pri}(\k_2)\rangle =
(2\pi)^3P_{\rm pri}(k_1)\delta^{(3)}(\k_1+\k_2)$. 
Note that $P_{\rm pri}(k)=A\,k^{n_s}$ is time-independent; all of the
time dependence of the power spectrum is contained in the transfer
function.  We calculate the time-dependent CDM and baryon transfer
functions using CAMB \citep{CAMB}.  
Figure \ref{psi} illustrates these velocity correlations. On small 
scales, velocities at nearby points are almost perfectly correlated,
$\psi_1\simeq 1$ and $\psi_2\approx 0$. Towards larger scales of
order the sound horizon $\sim 150$ Mpc, the radial component of
the velocity is anti-correlated, since the relative CDM-baryon
velocities are sourced by structures on these scales. The
correlations fall off steeply on scales much larger than the sound
horizon. 

\begin{figure}
\centerline{\includegraphics[width=0.45\textwidth]{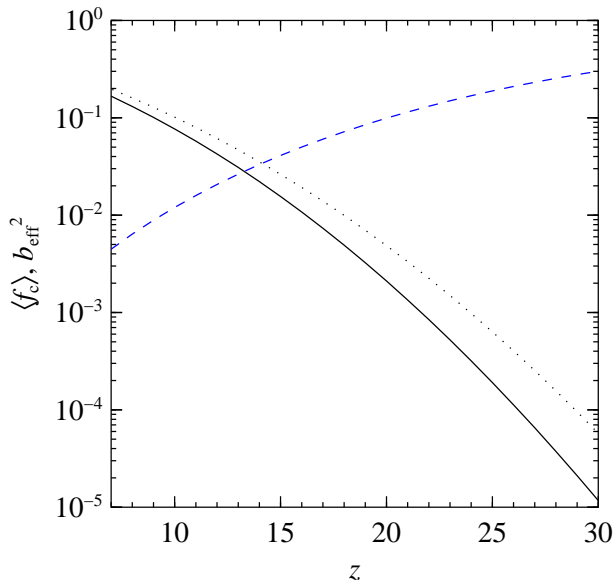}}
\caption{Redshift dependence of the mean collapse fraction 
$\langle f_c\rangle$ (solid black curve), the collapse fraction in
the absence of relative baryon-DM motions $f_c(v=0)$ (dotted black
curve), and the square of the
effective bias, $b_{\rm eff}^2$ (dashed blue curve), defined in
Eqn.\ (\ref{xif_approx}).
\label{fig:fc}
}
\end{figure}

Since we have assumed that the collapse fraction $f_c$ is a function
of the local (relative) velocity \v, we can easily compute the
probability distribution of $f_c$ and its moments by integrating over
the Gaussian distribution of \v. For example, the mean collapse
fraction is 
\begin{equation}
\langle f_c\rangle = \int P(\v) f_c(v) d^3\v =
\sqrt{\frac{2}{\pi}}\int f_c(v) e^{-v^2/2\sigma_1^2} \frac{v^2 dv}{\sigma_1^3}.
\end{equation}
We plot the redshift dependence of $\langle f_c\rangle$ in Fig.\
\ref{fig:fc}. 

Similarly, the two-point correlation function of $f_c$ is
\begin{equation}
\langle f_c(x) f_c(x+r)\rangle = 
\int d^3\v_1 d^3\v_2 P(\v_1,\v_2) f_c(v_1) f_c(v_2)
\end{equation}
where $\v_1$ and $\v_2$ are the velocities at points $\bm{x}$ and
$\bm{x}+\bm{r}$, respectively, whose probability $P$ is a Gaussian
distribution with covariance given by Eqn.\ (\ref{twopt}). This is a
6-D integral, however two of the integrals are elementary. If we
write $\bm{u}=\v/\sigma_1$ with components $u_r$ and $u_t$ parallel
and perpendicular to the separation vector, then
\begin{eqnarray}
\langle f_c(x) f_c(x+r)\rangle &=& 
\int\frac{du_{1r}du_{2r}du_{1t}du_{2t}}{2\pi(1-\psi_1^2)(1-\psi_r^2)^{1/2}}
f_c(v_1) f_c(v_2) 
u_{1t}u_{2t}I_0\left(\frac{\psi_1u_{1t}u_{2t}}{1-\psi_1^2}\right)
\nonumber \\
&& ~~ \times \exp\left(-\frac{1}{2}\left[\frac{u_{1t}^2+u_{2t}^2}{1-\psi_1^2}
+\frac{u_{1r}^2+u_{2r}^2-2\psi_ru_{1r}u_{2r}}{1-\psi_r^2}\right]\right),
\label{xif}
\end{eqnarray}
where $\psi_r=\psi_1+\psi_2$ and $I_0(x)$ is a modified Bessel
function. In various regimes, this integral may be simplified by
Taylor expansion. For example, at large radius where $|\psi_1|\ll 1$
and $|\psi_r|\ll 1$, the two-point function is approximately
\begin{eqnarray}
\langle f_c(x) f_c(x+r)\rangle &\approx& \frac{2}{\pi}
\int du_1 du_2 u_1^2u_2^2 f_c(v_1)f_c(v_2) e^{-(u_1^2+u_2^2)/2} \\
&& ~~\times \left[1+\left(\psi_1^2+\frac{\psi_r^2}{2}\right)
\left(1-\frac{u_1^2+u_2^2}{3} + \frac{u_1^2u_2^2}{9}\right)\right].\nonumber
\end{eqnarray}
Writing $1+\xi_f(r) = \langle f_c(x) f_c(x+r)\rangle/\langle f_c\rangle^2$,
we find
\begin{equation}
\xi_f(r) \simeq \left(\psi_1^2+\frac{\psi_r^2}{2}\right)
\left(1-\frac{\langle v^2f_c\rangle}{\sigma^2\langle f_c\rangle}\right)^2
\equiv b_{\rm eff}^2 \left(\psi_1^2+\frac{\psi_r^2}{2}\right),
\label{xif_approx}
\end{equation}
where $\sigma^2=\langle v^2\rangle = 3\sigma_1^2$ is the 3-D velocity
dispersion. 

This expression has a straightforward interpretation. We have assumed
that the collapse fraction is a function of the local velocity $v$.
Therefore, on large scales, $f_c$ will be a biased tracer of $v^2$.
The two-point correlation function of $v^2$ is simply
$2\psi_1^2+\psi_r^2$. So Eqn.\ (\ref{xif_approx}) is not surprising:
$f_c$ is indeed a biased tracer of $v^2$, with a large-scale bias
coefficient 
$b_{\rm eff} = \langle v^2f_c\rangle/\sigma^2\langle f_c\rangle-1$.
The form of this result is therefore generic.  The numerical value of
the bias coefficient depends on the precise details of the baryon
content of low-mass halos, which we have assumed to be a step-function
at $M_c$ for simplicity.  A more realistic model will have a slightly
different bias coefficient, but the result that $\xi_f\propto\psi_v$
holds generically as long as the gas content of halos depends on the
local relative bulk velocity.

\begin{figure}
\centerline{
\includegraphics[width=0.45\textwidth]{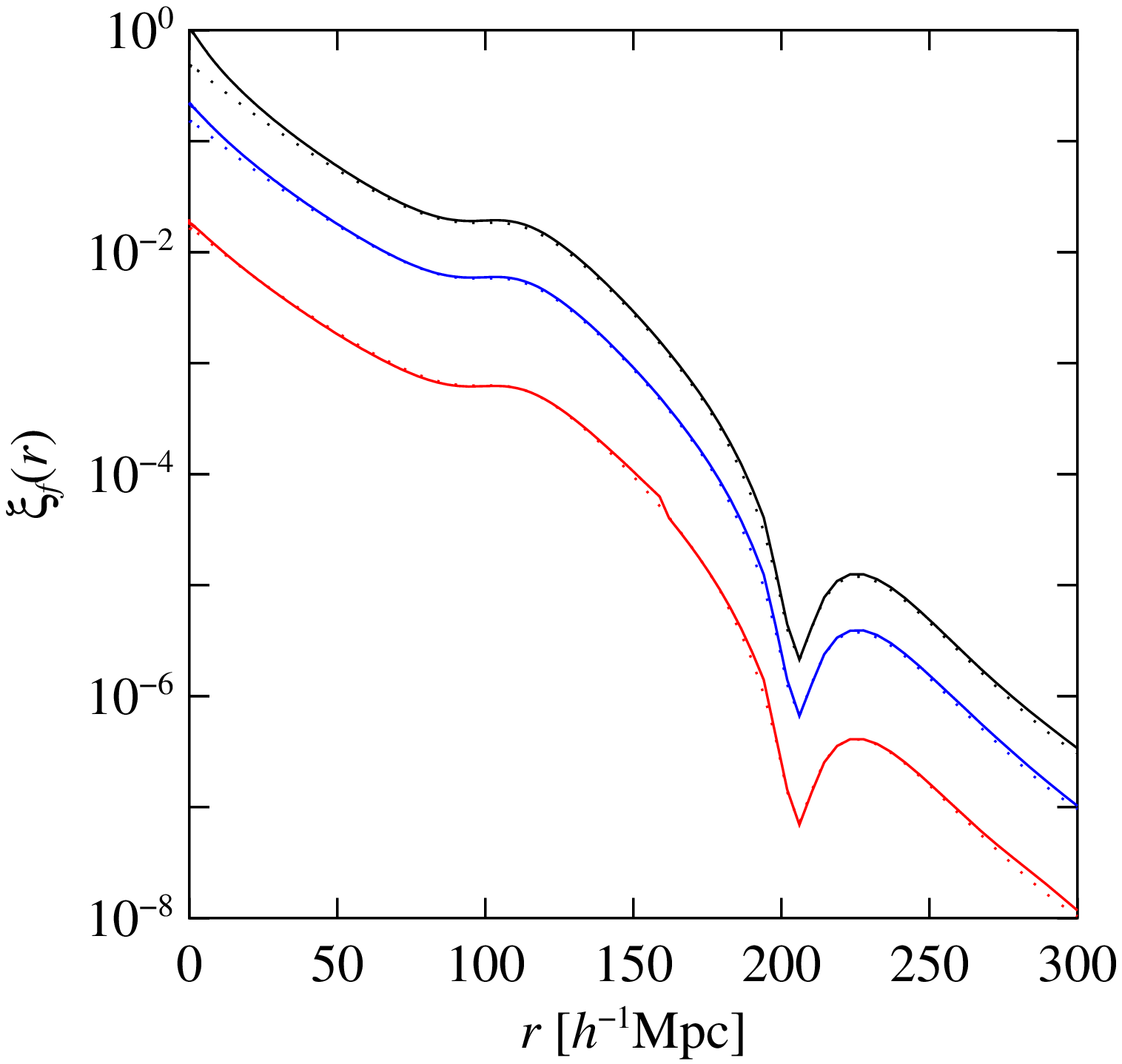}\quad
\includegraphics[width=0.45\textwidth]{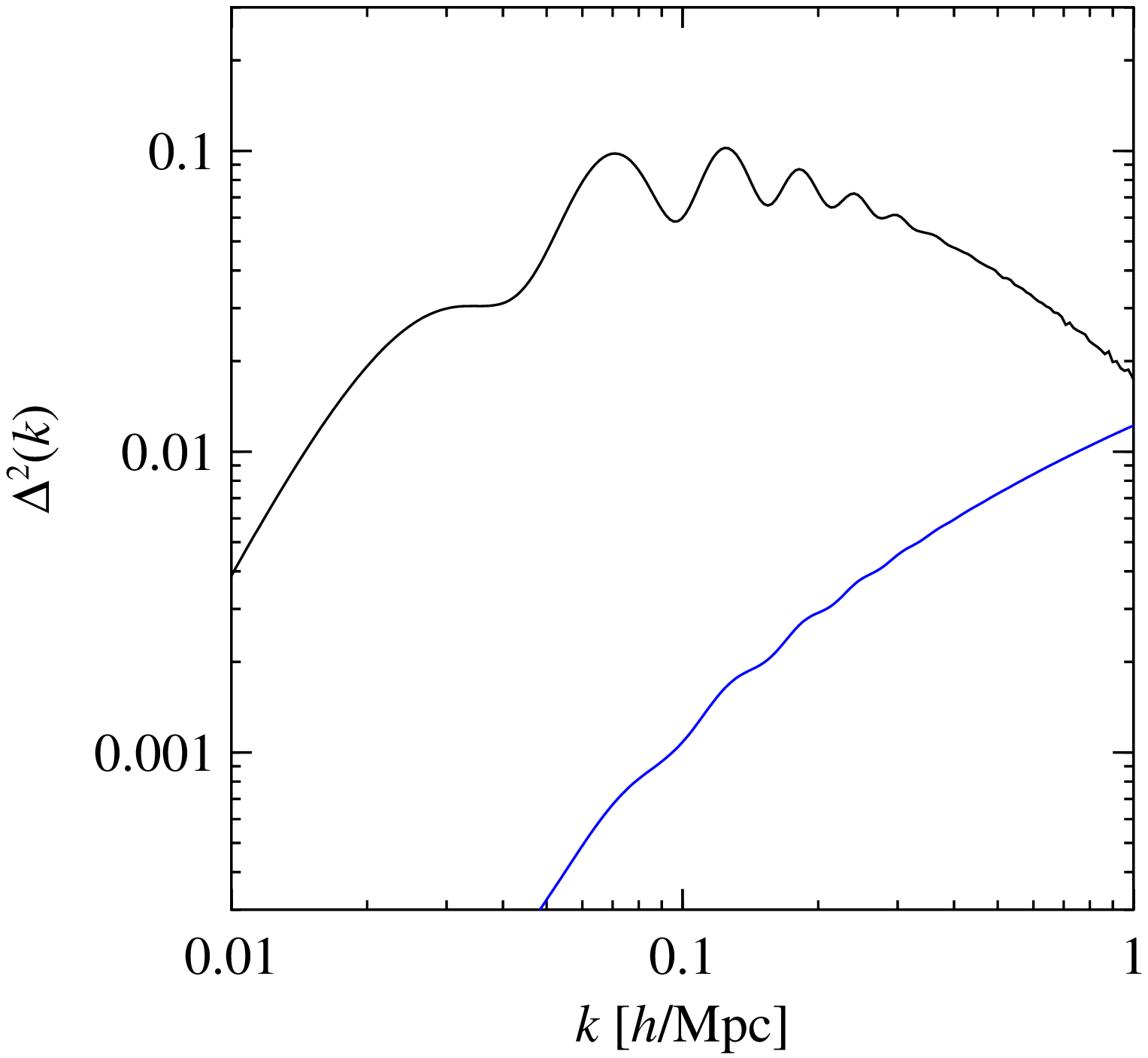}
}
\caption{(Left) The $f_c$ correlation function $\xi_f(r)$ at $z=10$
(red), $z=20$ (blue) and
$z=30$ (black). The solid curves show the full correlation function
calculated from Eqn.\ (\ref{xif}), while the dotted curves show the
linear bias approximation of Eqn.\ (\ref{xif_approx}). At small
radii, the linear bias approximation begins to break down, when
Poisson-like clustering can become significant. 
(Right) The dimensionless power spectrum $\Delta^2=k^3 P(k)/2\pi^2$ 
for fractional fluctuations in collapse fraction
$\delta_f=f_c/\langle f_c\rangle - 1$ (black) and
matter $\delta = \rho/{\bar\rho} - 1$ (blue) at $z=20$.
\label{fig:xi_f}}
\end{figure}

In Figure \ref{fig:xi_f} we plot the two-point correlation function and
power spectrum of the baryonic collapse fraction.  The two-point
function exhibits strong clustering on large scales, roughly $\sim
10\%$ fractional fluctuations on 100 Mpc scales at $z=20$.  There is
also a pronounced baryon acoustic oscillation feature, which is more
plainly visible in the power spectrum.  For comparison, we have also
plotted the matter power spectrum at the same redshift.  Two aspects
to note are that the collapse fraction has considerably enhanced power
on large scales, compared to the matter power spectrum, and that the
amplitude of the baryon oscillations are far larger in the $f_c$ power
spectrum.  As the figure illustrates, the baryon wiggles are typically
percent level in the matter $P(k)$, while in the $f_c$ power spectrum,
the baryon oscillations are order unity.

\section{Lyman $\alpha$ pumping}

The large-scale modulations of collapsed baryonic mass found in the
previous section can have several observable consequences. In this
section, we discuss one such consequence: the effect of relative
baryon-DM velocities on the 21-cm anisotropies preceding reionization.
The physics of 21-cm fluctuations at high redshift has been discussed
extensively in the literature; see \citet{Madau97} and \citet{Furl06} 
for reviews.

Prior to reionization, the gas kinetic temperature $T_{\rm kin}$ is
far below the CMB temperature: roughly speaking, 
$T_{\rm kin} \sim T_{\rm cmb}\times (1+z)/150$ for $z<100$. 
When the first stars begin to emit Lyman-$\alpha$ radiation, the
Wouthuysen-Field effect couples the spin temperature of neutral
hydrogen atoms to the gas kinetic temperature, allowing neutral gas to
be seen in 21-cm absorption against the cosmic microwave background.
Neglecting atomic collisions, the spin temperature is given by
\begin{equation}
T_s^{-1}=\frac{T_{\rm cmb}^{-1} + x_\alpha T_{\rm kin}^{-1}}{1+x_\alpha}
\label{spintemp}
\end{equation}
where the coupling coefficient $x_\alpha$ may be expressed in terms of
the ratio of Lyman $\alpha$ photons to H atoms, $n_\alpha$, as 
\citep{Furl06}
\begin{eqnarray}
x_\alpha &=& S_\alpha \frac{n_\alpha}{0.0767}\left(\frac{1+z}{20}\right)^2
\label{coupling} \\
S_\alpha &\approx& \exp\left(-0.803 T_{\rm kin}^{-2/3}
[\gamma\times10^6]^{-1/3}\right) \\
\gamma&=&\frac{H}{\lambda_\alpha \sigma_\alpha n_{\rm H\,I}}
\sim 3\times10^{-6} \frac{\Omega_m^{1/2}}{\Omega_b h (1+z)^{3/2}}. 
\end{eqnarray}
Given the spin temperature $T_s$, the optical depth to 21 cm
absorption is \citep{Madau97} 
\begin{equation}
\tau=\frac{3c^3 h n_{\rm H\,I} A_{10}}{32\pi H \nu_0^2 k_B T_s} \approx
0.155 h \frac{\Omega_b}{\Omega_m^{1/2}} (1+z)^{3/2} x_{\rm H\,I}
\left(\frac{T_s}{\rm K}\right)^{-1},
\label{tau}
\end{equation}
and the observed (i.e.\ redshifted) 21 cm brightness temperature
contrast against the CMB is then
\begin{equation}
\delta T_b = \frac{T_s-T_{\rm CMB}(z)}{1+z} (1-e^{-\tau}),
\label{contrast}
\end{equation}
where $T_{\rm CMB}(z) = 2.726 (1+z)$ K. Therefore, given the
Lyman-$\alpha$ intensity at any given point, we can compute the
observed temperature contrast.

We estimate the Lyman $\alpha$ intensity by assuming that, on average,
each collapsed baryon emits $N_\alpha$ Lyman $\alpha$ photons that
escape into the intergalactic medium. For simplicity, we assume that
photons are emitted with a flat spectrum, $\nu dN/d\nu=\,$const, but it
is straightforward to generalize our results for an arbitrary emission
spectrum. The local Lyman-$\alpha$ photon number density at each
point in space and redshift is then given by a convolution of the
collapsed baryon density with a retarded Green's function,
\begin{equation}
n_\alpha = N_\alpha \int d^3r\,d\eta \frac{df_c}{d\eta}(\bm{r},\eta) 
\frac{\delta(\eta+r/c)}{4\pi r^2 c},
\end{equation}
or in Fourier space,
\begin{equation}
n_\alpha(\bm{k},\eta_0) = N_\alpha \int_{0}^{\eta_0} d\eta
\frac{df_c}{d\eta}(\bm{k},\eta) j_0(kc(\eta_0-\eta))
\end{equation}
where $\eta$ is conformal time and $j_0(x)=\sin(x)/x$. 
We can further simplify this expression by using our previous result
that, on large scales, $f_c(\bm{k})$ has time dependence 
$\propto\langle f_c\rangle b_{\rm eff}=\langle f_c\rangle
-\langle v^2 f_c\rangle/\sigma^2$. Therefore, we have
\begin{equation}
n_\alpha(\bm{k},z) = N_\alpha f_c(\bm{k},z) W(k,z)
\label{intensity}
\end{equation}
where the smoothing filter is
\begin{equation}
W(k,z)=\frac{1}{\langle f_c\rangle b_{\rm eff}}
\int_{0}^{\eta_0} d\eta \frac{d\langle f_c\rangle b_{\rm eff}}{d\eta}
j_0(kc(\eta_0-\eta)).
\label{window}
\end{equation}
In this expression, $\eta_0$ is the conformal time at redshift $z$,
and note that $d/d\eta = -H\,d/dz$. 

The propagation of Lyman $\alpha$ photons over large distances
considerably damps the spatial fluctuations in Ly $\alpha$ intensity.  
However, our discussion so far has neglected an important effect:
Lyman $\alpha$ photons can only travel a limited distance. A rest
frame Lyman $\alpha$ photon which participates in pumping was emitted
at some distance bluewards of Lyman $\alpha$, redshifting as it
travels.  The higher the frequency at emission, the longer the
distance that the photon travels before redshifting into Lyman
$\alpha$.  However, a photon that is emitted at a wavelength shorter
than Lyman $\beta$ will be absorbed in the neutral intergalactic
medium, and ultimately lost to double photon decay, before it can
redshift into Lyman $\alpha$.  This means that 
gas clouds at redshift $z$ cannot be pumped by photons emitted by
sources at redshift $z_{\rm emit} > z_{\rm hor}$, where 
$(1+z_{\rm hor})=(32/27)\times(1+z)$.  This gives a natural maximal
horizon distance for Lyman $\alpha$ pumping.  
Conceivably, the propagation distance could be even shorter given
sufficient molecular opacity, either in the host minihalos or in the
intergalactic medium \citep{Ricotti01}, but we disregard this
possibility in our calculations.

If $z_{\rm hor}/z - 1 \ll 1$, we can approximate Eqn.\ (\ref{window})
as
\begin{equation}
W(k,z)\approx-\frac{1}{\langle f_c\rangle b_{\rm eff}}
\int_{z}^{z_{\rm hor}} dz_e \frac{d\langle f_c\rangle b_{\rm eff}}{dz}
j_0\left(\frac{kc}{H}(z_e-z)\right).
\label{window_approx}
\end{equation}
This shows how the shape of the smoothing window depends on the
formation history of collapsed baryonic objects and their Lyman
$\alpha$ emission.  As noted above, the shape of $W(k,z)$ also
depends on the spectrum of escaping UV emission from the first stars,
which will generally be much more complicated than we have assumed
here. Fortunately, it is entirely straightforward to compute how $W$
changes when realistic spectra and opacity are used instead of the
flat spectrum and sharp cutoff that we have adopted for simplicity.
Given this expression for the smoothing window, we can compute the
number of Lyman $\alpha$ photons per atom, $n_\alpha$, using Eqn.\
(\ref{intensity}), which then gives $x_\alpha$ and the spin
temperature $T_s$ using Eqns.\ (\ref{spintemp}-\ref{coupling}). Given
$T_s$, we compute the optical depth and brightness temperature using
Eqns.\ (\ref{tau}-\ref{contrast}).

\begin{figure}
\centerline{
\includegraphics[width=0.45\textwidth]{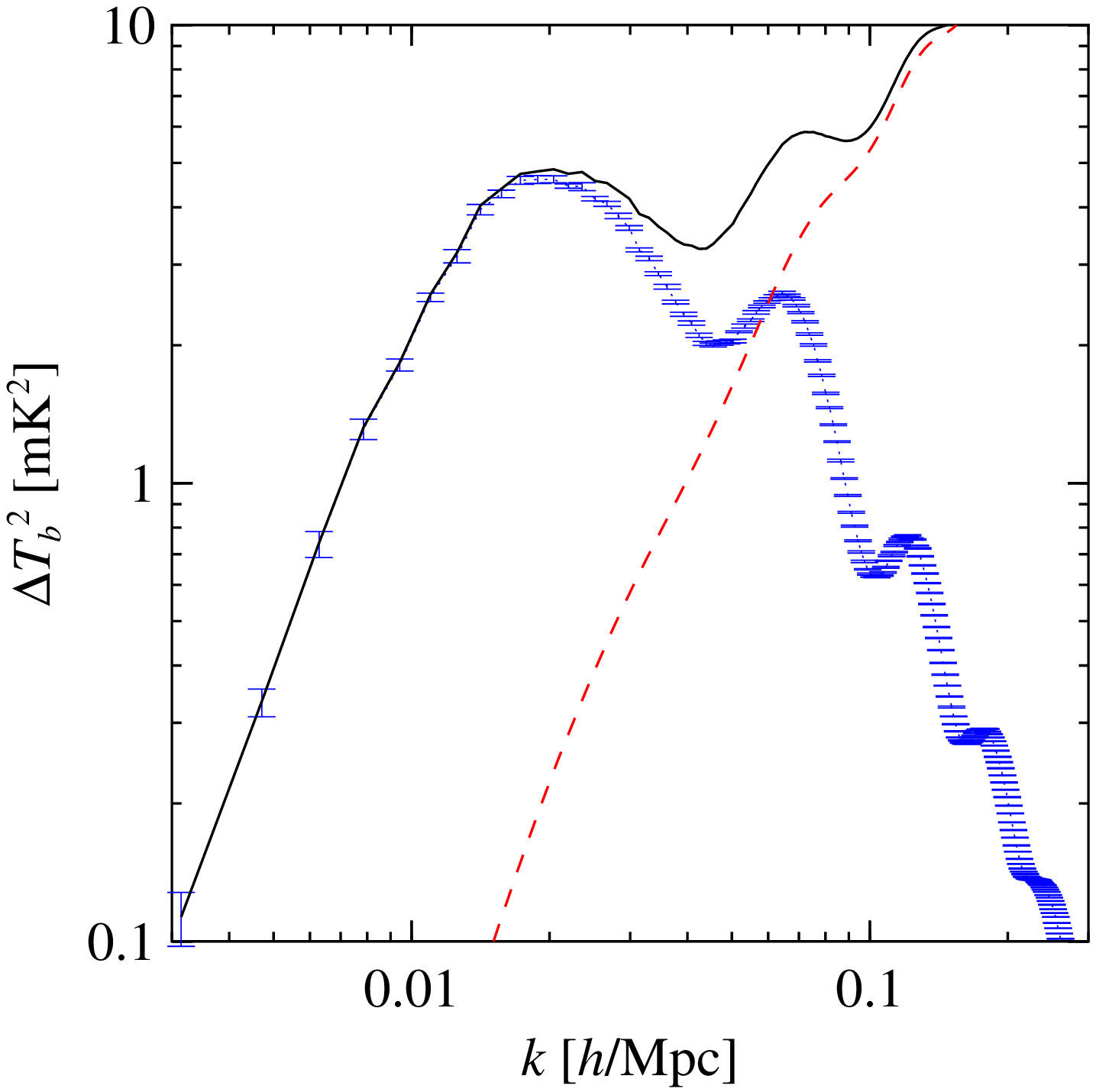}\quad
\includegraphics[width=0.45\textwidth]{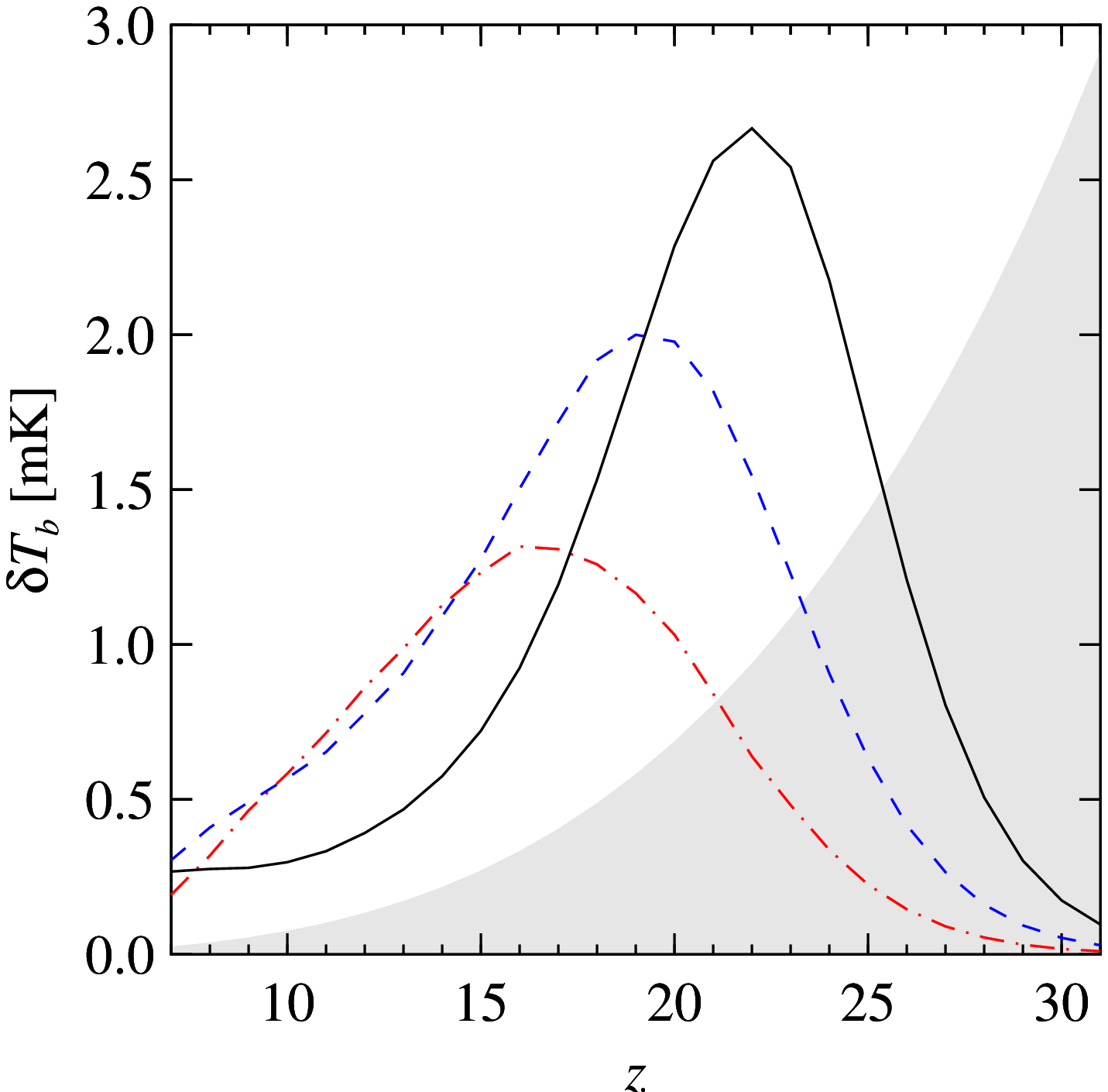}
}
\caption{
(Left) The 21 cm temperature power spectrum at redshift
$z=20$, for $N_\alpha=100$ Lyman $\alpha$ photons per collapsed atom.
The blue curve with error bars shows the brightness power spectrum
from UV sources that trace minihalos, while the red dashed curve
shows the fluctuations expected for uniform
Lyman $\alpha$ radiation that pumps gas with density fluctuations
tracing the matter density fluctuations at this redshift.  Since
density and velocity are uncorrelated, these two power spectra add
linearly (solid black curve). 
(Right) RMS temperature fluctuations $\delta T_b$ as a function of
redshift $z$ for a Gaussian beam with a diffraction limit of a 500m
dish (such as FAST). A 300m dish measures a slightly lower signal.
Different curves are: $N_\alpha = 10$ (red dot-dash), 30 (blue dashed)
and 100 (black solid).  The shaded area indicates the $1-\sigma$
errors per unit redshift expected for 1000 days of observations using
a 500m dish like FAST.  
\label{pumpfig}
}
\end{figure}

The Lyman-$\alpha$ intensity $n_\alpha$ is a linear function of
the collapse fraction, so its power spectrum is simply the product of
the $f_c$ power spectrum with the (square of the) window function,
Eqn.\ (\ref{window}). The brightness temperature is a nonlinear but
local function of $n_\alpha$. Therefore, on large scales it is a
biased tracer of the intensity field, and its power spectrum will be
proportional to the $n_\alpha$ power spectrum, with some
proportionality coefficient. We could write down an analytic
expression for this bias coefficient in terms of the $N$-point
correlation functions of $n_\alpha$, but it is simpler to calculate it
by simulation instead. Accordingly, we have generated realizations of
the brightness temperature field. We first generate realizations of
the Gaussian random relative velocity field $v_{cb}$, which we then
transform into collapse fraction $f_c$ using Eqns.\
(\ref{Mc}-\ref{eqn:fc}), replacing $c_s\to c_{s,{\rm eff}}$ as
described in \S\ref{sec:fc}. From the collapse fraction, we compute
the Lyman-$\alpha$ intensity using Eqn.\ (\ref{intensity}), which then
gives the spin temperature $T_s$, optical depth $\tau$ and brightness
temperature contrast $\delta T_b$ using Eqns.\
(\ref{spintemp}-\ref{contrast}). We generate realizations in a 2
$h^{-1}$Gpc box of $1024^3$ pixels at a variety of different
redshifts, varying the number of Lyman-$\alpha$ photons per collapsed
baryon, $N_\alpha$. 

Figure \ref{pumpfig} illustrates the brightness temperature power
spectrum. 
%We plot the angular power spectrum, calculated from the 3-D
%power spectrum using a flat sky approximation,
%\begin{equation}
%C_l \approx \frac{1}{\pi r} \int_{l/r}^\infty dk
%\frac{k P(k)}{\sqrt{k^2r^2 - l^2}}.
%\end{equation}
As expected, the $\delta T_b$ power spectrum is
proportional to the product of the $f_c$ power spectrum, multiplied by
the square of the window function $W(k,z)$ given by Eqn.\
(\ref{window}), which suppresses small-scale fluctuations in the
brightness temperature. The shape of the power spectrum is therefore
simple to calculate. The power spectrum peaks on the scale of the
Lyman-$\alpha$ horizon, and exhibits damped but pronounced acoustic
oscillations at higher wavevectors.

The amplitude of the power spectrum is
a nontrivial function of $N_\alpha$ and $z$. At high redshift, when
the collapse fraction is small and the Lyman-$\alpha$ pumping
intensity $n_\alpha$ is weak, the spin temperature is close to the CMB
temperature, with small fluctuations proportional to $n_\alpha$. The
brightness temperature contrast therefore grows rapidly in time.
Eventually, however, the spin temperature begins to saturate at the
gas kinetic temperature $T_{\rm kin}$. As $n_\alpha$ becomes very
large, the spin temperature begins to approach a uniform value
everywhere, $T_s \to T_{\rm kin}$, and so the brightness temperature
fluctuations actually diminish with increasing $n_\alpha$.
Accordingly, the $\delta T_b$ power spectrum peaks and then decreases
towards lower redshift. Of course, our assumption that $T_{\rm kin}$
remains close to its adiabatic value becomes suspect in this regime.
Eventually, a sufficiently strong Lyman-$\alpha$ intensity will not
merely pump the 21-cm transition, but will appreciably heat the gas as
well.

\section{Observational prospects}

In order to measure the very large scale structure, one needs high
brightness sensitivity on BAO scales. At $z>10$, the angular scale
of BAO changes little, and $\sim 10^\prime$ scales are important, just
as they are in the CMB.

To achieve high brightness sensitivity, a filled aperture is
desirable. Telescopes such as Arecibo and FAST \citep{FAST}
would be well suited.
With filled apertures of 300m and 500m respectively, their angular
resolution for 21 cm at $z\sim 20$ is 40 and 30 arc minutes,
respectively.  We focus our attention on FAST, which is expected to
observe at sufficiently long wavelengths.
At these low frequencies, sensitivities are sky limited,
with $T_{\rm sys} \sim 3000$K. It is also straightforward to observe
with a focal plane array, so we reference our forecasts to a 100 pixel
array, which would be a 40m dipole array. Such a focal plane array
would allow primary beam illumination to compensate with frequency to
make frequency independent beams, enabling accurate foreground
subtraction. The maximum transverse $k_\perp=0.08$.

We further assume the sky
is drift scanned, which minimizes systematic errors, and makes this
experiment comparable to the recent GBT 21cm intensity detection
\citep{Chang10}. For a square array, the field of view is 10 beam
width, or about 5 degrees. We use half the scanned area as useful,
allowing for galaxy and point source cuts.

At zenith for a latitude of 30 degrees, this scans 1000 square
degrees. We use an integration time of 1000 days on the sky. The
exposure time per pixel is $5\times 10^6$ seconds. In a bandwidth of 2 MHz,
this results in a pixel noise of 1 mK, well matched to the expected
signal. This map contains $\sim 10^5$ pixels, so one expects to
measure the power on the beam scale with $\sim 100 \sigma$.

The specific forecast parameters used for the noise estimate in figure
\ref{pumpfig} are: $T_{\rm sky}=300 (\nu/150 {\rm MHz})^{-2.7}$, 24
beams on the sky (the equivalent of a LOFAR station signal processor),
and a diffraction limited beam.

% d[z=20,0.25]=8.055/h Gpc, reduces by 21.55/h Mpc to z=19

\section{Discussion}

We have investigated the effect of relative motions between baryons
and dark matter on the formation of the smallest galaxies at high
redshift, $z\sim 20$. The formation of galaxies in minihalos of mass 
$M\lesssim 10^6 M_\odot$ is modulated by large-scale bulk velocities
between gas and dark matter, and so the clustering of these objects
will contain a contribution proportional to the relative velocity
two-point function. The velocity power spectrum exhibits significant
correlations on large scales of order 100 Mpc, with pronounced
baryon acoustic oscillations. Accordingly, the large-scale
clustering of minihalos exhibits similar features, and the large-scale
correlations of any observable that traces minihalos will show similar
behavior. We illustrated this in the previous section with a
calculation of the 21 cm absorption power spectrum prior to
reionization, when the kinetic temperature of intergalactic gas is
much colder than the CMB temperature.  Our predictions for 21 cm
absorption are model dependent, principally depending on the number of
Lyman $\alpha$ photons emitted per collapsed baryon.

\begin{figure}
\centerline{
\includegraphics[width=0.45\textwidth]{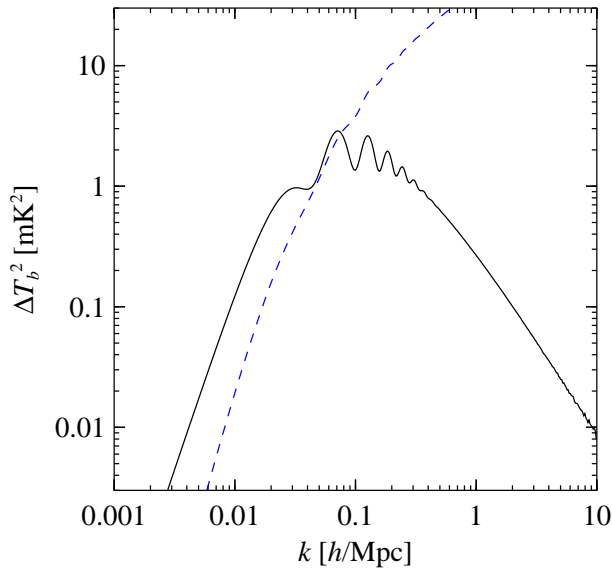}
}
\caption{The 21 cm temperature power spectrum 
$\Delta T_b^2 = k^3 P(k)/2\pi^2$ at $z=10$ (solid black),
assuming that the ionized fraction is proportional to the local
baryon collapsed fraction, $x_e\propto f_c$. For comparison, we
also plot the matter power spectrum multiplied by $(1+z)\times(9.4
{\rm mK})^2$, as the dotted blue curve.
\label{patchy}
}
\end{figure}

A similar argument holds for any observable that traces minihalos.
For example, suppose that minihalos make a significant contribution to
the ionizing flux at the time of reionization. Then the large-scale
power spectrum of patchy reionization will contain a contribution
proportional to the bulk velocity power spectrum, as can easily be
seen. During reionization, let us assume that $T_s \gg T_{\rm CMB}$, 
and that fluctuations in the ionized fraction $x_e$ trace the local
collapsed fraction $f_c$. In this regime, the observed temperature
contrast in Eqn. (\ref{contrast}) becomes independent of the spin
temperature 
\begin{equation}
\delta T_b \simeq 0.155 {\rm K}\ h \frac{\Omega_b}{\Omega_m^{1/2}}
(1+z)^{1/2} x_e \approx 9.4 {\rm mK}\, (1+z)^{1/2} x_e.
\end{equation}
We fix the proportionality constant relating $x_e$ and $f_c$ by
requiring that the Thomson scattering optical depth $\tau$ cannot
exceed the WMAP bound, $\tau=0.088$. We assume that reionization
completes at $z_{\rm re}=7$, and assume that $x_e$ at $z>z_{\rm re}$
is patchy and traces $f_c$. This provides an upper bound on the
patchy reionization signal from minihalos. The resulting power
spectrum is shown in Fig.\ \ref{patchy}. As expected, the minihalo
contribution to the power spectrum shows pronounced baryon acoustic
oscillations, which will be at least partially smoothed when we
account for the $\sim 5-10$ Mpc sizes of ionized bubbles during
reionization.

Although this signal could be present, detecting it during
reionization may be challenging in practice, since there are other
sources of large-scale power in 21-cm correlations. For example, the
clustering of the ionizing sources will reflect the clustering of the
halos hosting those sources, which trace the clustering of the matter
field. Because matter density fluctuations are uncorrelated with bulk
velocity fluctuations, these two terms simply add in the overall power
spectrum of the sources. Realistically, we expect the clustering from
bulk velocities to be subdominant to the clustering from density
perturbations over many of the scales of interest during reionization
(see Fig.\ \ref{patchy}). On the other hand, TH have shown that the
halo power spectrum itself also contains a term proportional to the
relative bulk velocity power spectrum, for halos of order $M\sim 10^6
M_\odot$.

Another place where these effects may be observable is in the CMB itself. 
At small angular scales, the so called kinetic Sunyaev-Zeldovich effect, 
caused by radial Doppler motions of electrons off which photons scatter, can be 
important \citep[e.g.][]{HH}. This effect is enhanced if reionization
is patchy \citep{McQuinn2005}.  
If these reionization patches have suppressed bulk velocities then the effect 
will be reduced. On the other hand, the large scale correlation of reionization 
patches predicted here will lead to large scale fluctuations in optical depth, 
inducing a modulated suppression of primary CMB fluctuations, 
which may be detectable with a higher order correlation analysis in CMB.

At low redshifts, $z\lesssim 5$, the typical masses of collapsing
halos are generally much larger than minihalo masses, and so we
would naively expect the effects of relative velocities between DM and
baryons to become unimportant.  However, in principle the signatures
of minihalos could persist even in late-time observables.  For
example, if reionization is patchy on $\sim 100$ Mpc scales due to
minihalos, the subsequent star formation history inside patches that
reionize early could differ from patches that reionize later.  This
could lead to spatial variations in galaxy formation at late times, on
scales of order the BAO scale.  This is potentially worrisome for BAO
probes of dark energy \citep[e.g.][]{BOSS}, which rely on precise
determination of the BAO feature in the galaxy two-point function.
Any contamination from minihalo effects could shift the location of
the BAO feature and thereby create a bias in measurements of the
equation of state parameter $w$, analogous to the results of
\citet{Pritchard07}. 
Removing this source of systematic uncertainty appears daunting.
\citet{Eisenstein07} have shown that BAO probes of dark
energy are quite insensitive to smooth distortions to the shape of the
power spectrum.  However, the minihalo effect could be far more
pernicious, since the velocity power spectrum is not smooth, but has
pronounced baryon oscillations that are presumably out of phase with
the oscillations in the matter power spectrum.  Marginalizing over an
unknown minihalo contaminant could significantly degrade BAO
constraints on dark energy.  This underscores the need for
more work on this subject.

In summary, the supersonic relative motions between baryons and dark
matter can dramatically affect the formation of the earliest collapsed
baryonic objects in the Universe.  We expect the abundance of
minihalos of mass $M\sim 10^5 M_\odot$ to be modulated on $\sim 100$
Mpc scales, and any tracer of minihalos should show the same
modulations, with highly pronounced baryon acoustic oscillations.
Specific predictions for any observable are necessarily model
dependent, but for plausible scenarios this large scale signal is
within the sensitivity range of existing and upcoming observatories.
This effect could allow us to detect the signature of the earliest
galaxies, written across the sky.

%\section*{Acknowledgments}
\acknowledgments{
We thank Chris Hirata for useful discussions.
The calculations presented in this paper have made use of publicly
available software, including CAMB\footnote{\tt http://camb.info},
CUBA\footnote{\tt http://www.feynarts.de/cuba},
FFTLog\footnote{\tt http://casa.colorado.edu/$\sim$ajsh/FFTLog} and 
GSL\footnote{\tt http://www.gnu.org/software/gsl}.  We thank the
authors of these libraries for making their software public.
This work is supported by the Swiss National Foundation
under contract 200021-116696/1 and WCU grant R32-2009-000-10130-0.
}

\newcommand{\jcap}{JCAP}

\end{document}